\documentclass[a4paper]{article}
\usepackage[psamsfonts]{amssymb}
\usepackage{amsmath}
\usepackage{cite}
\usepackage{epsfig}
\usepackage{graphicx}% Include figure files
\date{}

\author{M. Alimohammadi\footnote{alimohmd@ut.ac.ir}\ \ and A. Ghalee\footnote{am80gh@khayam.ut.ac.ir}
\\ {\small Department of Physics, University of Tehran,}
\\ {\small North Karegar Ave., Tehran, Iran.}}
\title{The phase-space of generalized Gauss-Bonnet dark energy }
\begin{document}
\maketitle
\begin{abstract}
The generalized Gauss-Bonnet theory, introduced by Lagrangian
$F(R,G)$, has been considered as a general modified gravity for
explanation of the dark energy. $G$ is the Gauss-Bonnet invariant.
For this model, we seek the situations under which the late-time
behavior of the theory is the de-Sitter space-time. This is done
by studying the two-dimensional phase space of this theory, i.e.
the $R-H$ plane. By obtaining the conditions under which the
de-Sitter space-time is the stable attractor of this theory,
several aspects of this problem have been investigated. It has
been shown that there exist at least two classes of stable
attractors: the singularities of the $F(R,G)$, and the cases in
which the model has a critical curve, instead of critical points.
This curve is $R=12H^2$ in $R-H$ plane. Several examples,
including their numerical calculations, have been discussed.

\end{abstract}
\section{Introduction}
Based on various observations, it is believed that our universe is
now in an accelerating phase~\cite{ries}. Although the origin of
this accelerated expansion is not yet known, almost all data
indicate that nearly 70$\%$ of the present universe is composed of
dark energy, the physical object that induces the negative
pressure.

There are two main classes of models that have been introduced as
candidates of dark energy. The first class is based on the
Einstein cosmology but with extra physical object as the source of
dark energy. The scalar field (one-component or multicomponents)
models~\cite{ratra}, the scalar-tensor theories~\cite{padm} and
the k-essence models~\cite{chiba} are examples in this context.

The second class of the models is based on the assumption that the
gravity is being (nowadays) modified. The simplest one is obtained
by adding a cosmological constant term to Einstein action. This
model suffers two important problems known as cosmological
constant and coincidence problems~\cite{wei}. Also, the
cosmological constant model is a static model of dark energy and
has not any dynamical behavior. The other modified gravity models
are those that are based on the new actions.

The first family of these modified gravity theories are those
known as $f(R)$ gravity, with the action
\begin{equation}\label{1}
S=\int {\rm d}^{4}x\sqrt{-g}{}\hspace{1ex}\left[
\frac{1}{2\kappa^{2}}f(R)+{\cal L}_m \right].
\end{equation}
In $\hbar =c=G=1$ units, $\kappa^{2} = 8\pi $, $R$ is the Ricci
scalar and ${\cal L}_m$ is the Lagrangian density of dust-like
matter. Many features of $f(R)$ gravity models, such as local
gravity tests, have been studied~\cite{hu}.

Another well motivated curvature invariant, beyond the Ricci
scalar, is the Gauss-Bonnet (GB) term
\begin{equation}\label{2}
G=R^{2}-4R_{\mu\nu}R^{\mu\nu}+R_{\mu\nu\xi\sigma}R^{\mu\nu\xi\sigma},
\end{equation}
which is inspired by string theory~\cite{free} and is a
topological invariant in four dimensions. The second family of
modified gravity theories is known as $f(G)$ gravity and is
defined through
\begin{equation}\label{3}
S=\int {\rm d}^{4}x\sqrt{-g}{}\hspace{1ex}\left[
\frac{1}{2\kappa^{2}}R+f(G)+{\cal L}_m \right].
\end{equation}
This model has gained special interest in cosmology \cite{sany}
and its coupling to scalar fields, as it naturally appears in
low-energy string effective actions~\cite{free}, can introduce
extra dynamics to this model. Other aspects of modified GB
gravity, such as the possibility of describing the inflationary
era, transition from deceleration phase to acceleration phase,
crossing the phantom-divide-line, and passing the solar system
test have been discussed in \cite{noi}.

The natural generalization of action ({\ref{3}) is the generalized
Gauss-Bonnet dark energy, which have been introduced in
\cite{free,koi} :
\begin{equation}\label{4}
S=\int {\rm d}^{4}x\sqrt{-g}{}\hspace{1ex}\left[ F(R,G)+{\cal L}_m
\right].
\end{equation}
Clearly the $f(R)$ gravity and $f(G)$ gravity are special examples
of modified $F(R,G)$ gravity. The hierarchy problem of particle
physics and the late-time cosmology have been studied in $F(R,G)$
models \cite{km}. Recently, the behavior of these models in
phantom divide line crossing and deceleration to acceleration
transition, including the contribution of quantum effects to those
phenomena, have been studied in \cite{ali}. It has been shown that
the quantum effects can induce these transitions, when they are
classically forbidden.

One of the important characteristics of all dynamical systems,
including the dynamical models of dark energy, is their late-time
behaviors, studied in a framework known as attractor solution of
dynamical systems, which has been deeply studied in mathematics.
For dark energy models, the attractor solutions of scalar theories
and some of the modified gravity theories have been studied in
\cite{alim,fn,bar}.

The main step in studying the attractor solution of a dynamical
system is considering a set of suitable dynamical variables
$x_1(t),\dotsc,x_n(t)$, such that their first time derivatives
${\rm d}x_i/{\rm d}t$ do not depend explicitly on time:
\begin{equation}\begin{split}\label{5}
&\frac{{\rm d}x_1}{{\rm
d}t}=F_1(x_1,\dotsc,x_n),\\
&\hspace{.8cm}\vdots\\
&\frac{{\rm d}x_n}{{\rm d}t}=F_n(x_1,\dotsc,x_n).
\end{split}\end{equation}
The space constructed by variables $x_1,\dotsc,x_n$ is called the
phase space of the system and the system of equations ({\ref{5})
is said to be autonomous. The functions $x_1(t),\dotsc,x_n(t)$
define a path in the phase space and there is a unique path which
passes any specific initial values
$x_1(t_0)=x_{1,0},\dotsc,x_n(t_0)=x_{n,0}$, i.e. the paths do not
intersect one another. The only exception to this statement occur
at points $(x_{1,c},\dotsc,x_{n,c})$ where
\begin{equation}\label{6}
F_1(x_{1,c},\dots,x_{n,c})=0,\,\dots\,
,F_n(x_{1,c},\ldots,x_{n,c})=0.
\end{equation}
These points are called critical points, and any paths near these
points, under specific conditions, will lead them at $t\rightarrow
\infty$. In these cases, the critical points are called the stable
attractors.

The present paper is devoted to the study of the phase space and
attractor solutions of generalized Gauss-Bonnet dark energy
models. We will consider the $R-H$ space ($R$ is the Ricci scalar
and $H$ is the Hubble parameter) as the phase space of these
models and show that in special case of $F(R,G)=f(R)$, the results
of \cite{far}, in which some features of phase space of $f(R)$
have been studied, are reproduced. The choice of this phase space,
which is the only possible choice in general $F(R,G)$ model, has
an important property. Since the attractor solutions are those
which asymptotically lead to $\dot{R}=0$ and $\dot{H}=0$, or
$R=R_c$ and $H=H_c$ ($R_c$ and $H_c$ are some constant values),
our phase-space study is in fact the study of possible de-Sitter
solutions of modified generalized GB gravity.

The scheme of the paper is as follows. In section 2, the set of
autonomous equations of $F(R,G)$ models is obtained and the
condition of existence of stable attractors is discussed. Some
specific examples of $F(R,G)$ models which admit the stable
attractors are investigated in section 3, and it is shown that the
numerical studies confirm our results. Section 4 is devoted to the
cases where the standard linear approximation method, used in
obtaining the stability behavior of solutions, does not work. In
section 5, it is shown that the singular points of the Lagrangian
are always the stable attractors, and finally in section 6, the
interesting cases where the critical points replaced by critical
curves are studied. It is shown that all these critical curves
always behave as stable attractor curves. We end the paper with a
conclusion in section 7.

\section{Critical points of $F(R,G)$ gravity}

Consider the generalized GB dark energy model with action
(\ref{4}). Variation of this action with respect to the metric
$g_{\mu\nu}$ results in ~\cite{km}
\begin{equation}\begin{split}\label{7}
&\frac{1}{2}g^{\mu\nu}F(R,G)-2F_{G}(R,G)RR^{\mu\nu}
+4F_{G}(R,G)R^{\mu}_{~\rho}R^{\nu\rho}\\
&-2F_{G}(R,G)R^{\mu\rho\sigma\tau}R^{\nu}_{~\rho\sigma\tau}
-4F_{G}(R,G)R^{\mu\rho\sigma\nu}R_{\rho\sigma}
+2(\nabla^{\mu}\nabla^{\nu}F_{G}(R,G))
R\\&-2g^{\mu\nu}(\nabla^{2}F_{G}(R,G))
R-4(\nabla_{\rho}\nabla^{\mu}F_{G}(R,G))
R^{\nu\rho}-4(\nabla_{\rho}\nabla^{\nu}F_{G}(R,G))R^{\mu\rho}\\
&+4(\nabla^{2}F_{G}(R,G))R^{\mu\nu}
+4g^{\mu\nu}(\nabla_{\rho}\nabla_{\sigma}F_{G}(R,G))R^{\rho\sigma}
-4(\nabla_{\rho}\nabla_{\sigma}F_{G}(R,G))R^{\mu\rho\nu\sigma}\\
&-F_{R}(R,G)R^{\mu\nu}+\nabla^{\mu}\nabla^{\nu}F_{R}(R,G)-
g^{\mu\nu}\nabla^{2}F_{R}(R,G)=0 .
\end{split}\end{equation}
Here, for simplicity, we do not consider the background matter
field, i.e.  ${\cal L}_m=0$. In equation (\ref{7}), $F_{R}$ and
$F_{G}$ are defined as following :
\begin{equation}\label{8}
F_{R}(R,G) = \frac{\partial F(R,G)}{\partial R}\hspace{4ex} ,
\hspace{4ex}F_{G}(R,G) = \frac{\partial F(R,G)}{\partial G} .
\end{equation}
For background metric, we consider, as usual, the spatially flat
Friedmann-Robertson-Walker (FRW) metric in co-moving coordinates
$(t,x,y,z)$ as follows:
\begin{equation}\label{9}
{\rm d}s^{2}=-{\rm d}t^{2}+a^2(t)({\rm d}x^{2}+{\rm d}y^{2}+{\rm
d}z^{2}) ,
\end{equation}
in which $a(t)$ is the scale factor. The $(t,t)$-component of
evolution equation (\ref{7}) then becomes
\begin{equation}\begin{split}\label{10}
 -6H^{2}F_{R}(R,G)=&F(R,G)-RF_{R}(R,G)+6H\dot{F}_{R}(R,G)\\
&+24H^{3}\dot{F}_{G}(R,G)-GF_{G}(R,G) .
\end{split}\end{equation}
$H=\dot{a}(t)/a(t)$ is the Hubble parameter. For this metric, the
Ricci scalar $R$ and the Gauss-Bonnet invariant $G$ are
\begin{equation}\label{11}
R=6(\dot{H}+2H^{2}),
\end{equation}
and
\begin{equation}\label{12}
 G=24H^{2}(\dot{H}+H^{2}),
\end{equation}
respectively. The equations (\ref{10})-(\ref{12}) are the
Friedmann equations of $F(R,G)$ gravity. The sum of $(i,i)$
components of eq.(\ref{7}) is obtained by using the time
derivative of eq.(\ref{10}) and the eqs.(\ref{11}) and (\ref{12}).

From eq.(\ref{11}), one has
\begin{equation}\label{13}
\dot{H}=\frac{R}{6}-2H^{2},
\end{equation}
which can be used to express $G$ from eq. (\ref{12}) as follows:
\begin{equation}\label{14}
G=4H^2(R-6H^{2}) ,
\end{equation}
from which
\begin{equation}\label{15}
\dot{G}=\frac{4}{3}HR^2+192H^5-32RH^3+4H^2\dot{R} .
\end{equation}
Using
\begin{equation}\label{16}
\frac{{\rm d}}{{\rm d}t}f(R,G) = f_R\dot{R}+f_G\dot{G},
\end{equation}
and eq.(\ref{15}), $\dot{R}$ can be extracted from eq.(\ref{10})
as follows:
\begin{equation}\label{17}
\dot{R} =
\frac{(R-6H^2)F_R+GF_G-F-288H^2({R}/{6}-2H^2)^2(F_{RG}+4H^2F_{GG})
}{6H(F_{RR}+8H^2F_{RG}+16H^4F_{GG})},
\end{equation}
\begin{equation}\label{18}
\dot{H}=\frac{R}{6}-2H^{2},
\end{equation}
where the second equation is the same as eq.(\ref{13}). The set of
above equations are the autonomous equations of $F(R,G)$ gravity.
The phase space of this problem is the two-dimensional ($R-H$)
space. In the right-hand side of eq.(\ref{17}), the expression
(\ref{14}) must be used for the Gauss-Bonnet invariant $G$.
Therefore the above equations are in the form
\begin{equation}\begin{split}\label{19}
&\dot{H}=f_1(R,H),\\
&\dot{R}=f_2(R,H) .
\end{split}\end{equation}
The critical points are found by setting the eqs.(\ref{17}) and
(\ref{18}) equal to zero. The result is
\begin{equation}\label{20}
\frac{1}{2}RF_R+GF_G-F=0 ,
\end{equation}
\begin{equation}\label{21}
R = 12H^2 .
\end{equation}
Equation (\ref{14}) also results in
\begin{equation}\label{22}
G = 24H^4={\frac {R^2}{6}}
\end{equation}
at critical points. In obtaining eq.(\ref{20}), it is assumed that
the denominator of eq.(\ref{17}) has finite value at critical
points. We will return to this assumption in section 5. Note that
in eq.(\ref{20}), $G$ must be replaced by eq.(\ref{22}). In the
case of $f(R)$ gravity, i.e. $F(R,G) = f(R)/2\kappa^{2}$, the
eqs.(\ref{17})-(\ref{21}) are reduced to the corresponding
relations in ref.~\cite{bar}.

Since the effective equation of state parameter is defined through
\begin{equation}\label{23}
\omega_{\rm{eff}}=\frac{p}{\rho}=-1-\frac{2}{3}\frac{\dot{H}}{H^{2}},
\end{equation}
at critical points where $\dot{H}=0$, one has
\begin{equation}\label{24}
\omega_{\rm{eff}}\rightarrow\omega_{c}=-1
\end{equation}
which is a characteristic of de-Sitter space-time.

To study the stability of each critical point, one must evaluate
the eigenvalues of matrix
\begin{equation}\label{25}
M=\left(\begin{array}{ll}\partial{f_1}/\partial{H}&\partial{f_1}/\partial{R}\\
\partial{f_2}/\partial{H}&\partial{f_2}/\partial{R}
\end{array}\right)_{R=R_c,H=H_c}.\
\end{equation}
$R_c$ and $H_c$ denote the values of $R$ and $H$ at the considered
critical point. The critical point is a stable attractor only when
the real parts of all the eigenvalues of matrix $M$ are negative.
For negative real eigenvalues, the stable critical point is called
a node, and for complex eigenvalues with negative real parts, the
stable attractor is called a spiral.

For autonomous equations (\ref{17}) and (\ref{18}), the matrix $M$
becomes
\begin{equation}\label{26}
M=\left(\begin{array}{llll} -4H&1/6\\
-2F_R/A&H
\end{array}\right)_{R=R_c,H=H_c},\
\end{equation}
where
\begin{equation}\label{27}
A=F_{RR}+8H^2F_{RG}+16H^4F_{GG} .
\end{equation}
Therefore the eigenvalues are
\begin{equation}\label{28}
\lambda_{1,2}=\frac{1}{2}\left[-3H\pm\sqrt{(3H)^2-4\left(\frac{F_R}{3A}-4H^2\right)}
\hspace{.1cm}\right]_{R=R_c,H=H_c}.\
\end{equation}
It is clear that the real part of eigenvalues $\lambda_{1}$ and
$\lambda_{2}$ are negative, if and only if
\begin{equation}\label{29}
\eta=\frac{F_R}{3A}-4H^2\biggm|_{R=R_c,H=H_c}>0.\
\end{equation}
This is the condition of stability of the attractors of $F(R,G)$
gravity. The attractors are node if $(3H_c)^2>4\eta$ and are
spiral if $(3H_c)^2<4\eta$ .

An interesting observation is that for $R$-independent Lagrangian
\begin{equation}\label{30}
F(R,G)=F(G),
\end{equation}
$\eta=-4H_{c}^2<0$ and therefore $\lambda_1>0$. So all the
critical points of these models are unstable.

\section{Some examples of stable attractors}
In this section we will discuss two classes of $F(R,G)$ models
which lead to stable attractors.

\subsection{$F(R,G)=F(RG)$ models }
For these models, the equation (\ref{20}) results in
\begin{equation}\label{31}
\frac{3}{2}xF'(x)-F(x)=0,
\end{equation}
where $x=RG$ and $F'(x)=\frac{{\rm d}}{{\rm d}x}F(x)$. This
relation determines the critical values $x=x_c$. The stability
condition (\ref{29}) leads to
\begin{equation}\label{32}
\frac{F'(x)}{9xF''(x)+4F'(x)}-1>0
\end{equation}
at $x=x_c$. For $F(x)=x^n$ cases, it can be shown that
eq.(\ref{31}) does not have a nontrivial solution, except for a
very special case which we will discuss it later. For
$F(x)=x^n-c$, where c is a positive constant, eqs.(\ref{31}) and
(\ref{32}) result in:
\begin{equation}\label{33}
x_c=\left(\frac{c}{1-3n/2}\right)^{1/n},\
\end{equation}
and
\begin{equation}\label{34}
\frac{10}{18}<n<\frac{12}{18},\
\end{equation}
respectively. As an explicit example, we consider $n=11/18$ and
$c=1$:
\begin{equation}\label{35}
F(RG)=(RG)^{11/18}-1.\
\end{equation}
Eq.(\ref{33}) then results in
\begin{equation}\label{36}
x_c=R_c G_c=\frac{R_c^3}{6}=(12)^{18/11}.
\end{equation}
So
\begin{equation}\label{37}
(R_c,H_c)=(7.047,0.766).
\end{equation}
Numerical calculation of eqs.(\ref{10})-(\ref{12}) results in
Figs.(1)-(4) for phase space, $R(t),H(t)$, and $\omega(t)$
behaviors, respectively. These figures show that the point
(\ref{37}) is a stable attractor of spiral type.
\begin{figure}[hbt]\label{Figure_1}
\centering
\includegraphics*[height=5cm,width=5cm]{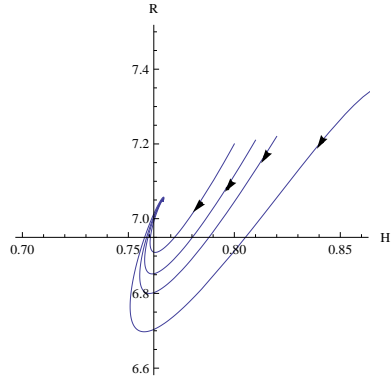}\hspace{1cm}
\caption{ The spiral paths in $R-H$ plane of
$F(R,G)=(RG)^{11/18}-1$
 model.}
\end{figure}
\begin{figure}[hbt]\label{Figure_2}
\centering
\includegraphics*[height=5cm,width=5cm]{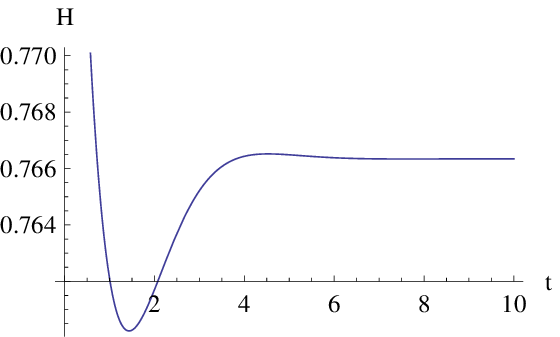}\hspace{1cm}
\caption{ The plot of $H(t)$ of $F(R,G)=(RG)^{11/18}-1$ model.}
\end{figure}
\begin{figure}[hbt]\label{Figure 3}
\centering
\includegraphics*[height=5cm,width=5cm]{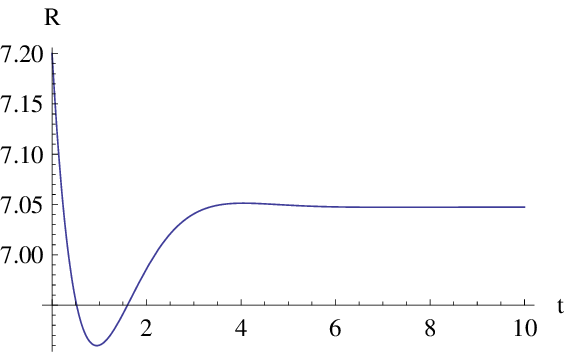}\hspace{1cm}
\caption{ The plot of $R(t)$ of $F(R,G)=(RG)^{11/18}-1$ model.}
\end{figure}
\begin{figure}[hbt]\label{Figure_4}
\centering
\includegraphics*[height=5cm,width=5cm]{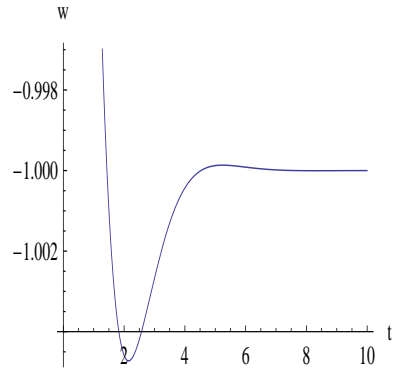}\hspace{1cm}
\caption{ The plot of $\omega(t)$ of $F(R,G)=(RG)^{11/18}-1$
model.}
\end{figure}

\subsection{$F(R,G)=R+Gf(R)$ models }
For this functional form of $F(R,G)$, the critical point equation
(\ref{20}), using eq.(\ref{22}), results in
\begin{equation}\label{38}
R^2f'(R)=6.
\end{equation}
This equation specifies $R_c$. The stability condition (\ref{29})
reduces to
\begin{equation}\label{39}
\frac{1}{R^3f''(R)/12+2}>1,\
\end{equation}
which must be calculated at $R=R_c$. For example for $f(R)=mR^n$
functions, where $m$ and $n$ are some constants, eqs.(\ref{38})
and (\ref{39}) result in
\begin{equation}\label{40}
R_c=\left(\frac{6}{mn}\right)^{1/(n+1)},\
\end{equation}
and
\begin{equation}\label{41}
-3<n<-1,
\end{equation}
respectively. Since $n$ is a negative number, $m$ must be chosen
negative so that $R_c$ in (\ref{40}) becomes a real positive
number. As a specific example, we consider $m=-1$ and $n=-2$, or
\begin{equation}\label{42}
F(R,G)=R-G/R^2.\
\end{equation}
$R_c$ and $H_c$ then become
\begin{equation}\label{43}
(R_c,H_c)=(\frac{1}{3},\frac{1}{6}).\
\end{equation}
Numerical results for Lagrangian (\ref{42}) are given by Figs.(5)
and (6), which are the paths in $R-H$ plane and $\omega(t)$,
respectively. Fig.(5) verifies that the critical point (\ref{43})
is a stable attractor.

\begin{figure}[hbt]\label{Figure 5}
\centering
\includegraphics*[height=5cm,width=5cm]{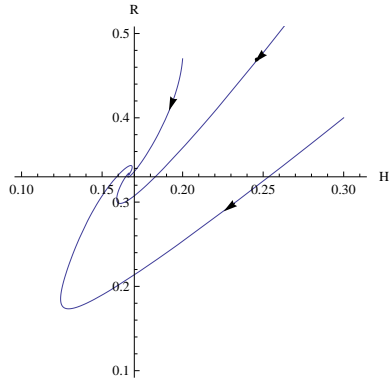}\hspace{1cm}
\caption{ The spiral paths of $F(R,G)=R-G/R^2$ model in $R-H$
plane. }
\end{figure}
\begin{figure}[hbt]\label{Figure_6}
\centering
\includegraphics*[height=5cm,width=5cm]{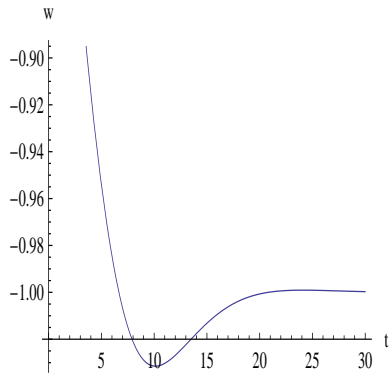}\hspace{1cm}
\caption{ The plot of $\omega(t)$ for $F(R,G)=R-G/R^2$ model. }
\end{figure}

\section{Critical points with zero eigenvalue }
As was mentioned previously, an attractor is stable if, and only
if, the real parts of all eigenvalues are negative. Now what we
can say if one, or more, of the eigenvalues are zero. The
appearance of zero eigenvalues may be rooted in nonindependency of
the chosen dynamical variables. For instance if we consider the
phase space of $F(R,G)$ models by three-dimensional $(R,H,G)$
space, one can show that besides the two eigenvalues $\lambda_1$
and $\lambda_2$ of eq.(\ref{28}), we have a third eigenvalue
$\lambda_3=0$. This indicates that we can reduce the
dimensionality of our phase space.

For the cases where the dimension of the phase space can not be
reduced, the appearance of zero eigenvalue means that the linear
approximation, which leads to the matrix (\ref{25}), is not
adequate and we must consider the higher order approximations to
determine the behavior of the critical points. See for example
~\cite{guh}.

In our $F(R,G)$ models, for the cases where
\begin{equation}\label{44}
\eta=\frac{F_R}{3A}-4H^2\biggm|_{R=R_c,H=H_c}=0,
\end{equation}
one has
\begin{equation}\label{45}
\lambda_1=0\hspace{.5cm},\hspace{.5cm}\lambda_2=-3H,
\end{equation}
and the higher order approximations must be used to determine
whether the considered critical point is stable or not. This is
done by a standard method which has been discussed, for example,
in~\cite{guh} .

At first, the eigenvectors of matrix $M$ for the eigenvalues
$\lambda_1=0$ and $\lambda_2=-3H$ must be calculated, which result
in
\begin{equation}\label{46}
\psi_1=\left(\begin{array}{ll} 1\\
24H_c
\end{array}\right)\hspace{.5cm},\hspace{.5cm} \psi_2=\left(\begin{array}{ll} 1\\
6H_c
\end{array}\right),
\end{equation}
respectively. Using the transformation matrix $T=(\psi_1,\psi_2)$,
the new phase-space basis $(U,V)$, i.e. the normal basis, can be
found from $(R,H)$ as follows
\begin{equation}\label{47}
\left(\begin{array}{ll} U\\
V\end{array}\right)=T^{-1}\left(\begin{array}{ll} R\\
H
\end{array}\right),
\end{equation}
which results in
\begin{equation}\begin{split}\label{48}
&U=\frac{1}{18H_c}(R-6H_cH),\\
&V=\frac{1}{18H_c}(24H_cH-R).
\end{split}\end{equation}
 To translate the critical point from $(R_c,H_c)$
to the origin of the phase space, we introduce $\mathcal{R}=R-R_c$
and $\mathcal{H}=H-H_c$. $\dot{\mathcal{R}}$ and
$\dot{\mathcal{H}}$, up to second order, then become
\begin{equation}\begin{split}\label{49}
&\dot{\mathcal{R}}=H_c\mathcal{R}-24H_c^{2}\mathcal{H}+D\mathcal{R}^2+B
\mathcal{H}^2+C\mathcal{R}\mathcal{H},\\
&\dot{\mathcal{H}}=\frac{1}{6}\mathcal{R}-4H_c\mathcal{H}-2\mathcal{H}^2.
\end{split}
\end{equation}
In above equations, the eqs.(\ref{17}) and (\ref{18}) have been
used and the coefficients $D$, $B$ and $C$ are defined by
\begin{equation}\label{50}
D=\frac{1}{2}\left(\frac{\partial^{2}{f_2(R,H)}}{\partial{R^{2}}}\right)\,\,\,
,\,\,\,
B=\frac{1}{2}\left(\frac{\partial^{2}{f_2(R,H)}}{\partial{H^{2}}}\right)\,\,\,
,\,\,\,
C=\left(\frac{\partial^{2}{f_2(R,H)}}{\partial{R}\partial{H}}\right).
\end{equation}
$f_2(R,G)$ is one introduced in eq.(\ref{19}) and all derivatives
are calculated at critical point ($R_c,H_c$). Note that the linear
terms in the right-hand-side of eq.(\ref{49}), result in the
matrix elements of $M$ in eq.(\ref{26}) for the case where
$\eta=0$, or $-2F_R/A=-24H^2$.

For the new phase space, introducing $\mathcal{U}=U-U_c$ and
$\mathcal{V}=V-V_c$, the eqs.(\ref{48}) and (\ref{49}) then result
in
\begin{equation}\label{51}
\dot{\mathcal{U}}=\left(\frac{2}{3}+\frac{B}{18H_c}\right)(\mathcal{U}+\mathcal{V})^{2}
+2DH_c(4\hspace{.1cm}\mathcal{U}+\mathcal{V})^{2}
+\frac{C}{3}(\mathcal{U}+\mathcal{V})(4\,\mathcal{U}+\mathcal{V})\hspace{.1cm},
\end{equation}
\begin{equation}\label{52}
\dot{\mathcal{V}}=-\left(\frac{8}{3}+\frac{B}{18H_c}\right)(\mathcal{U}+\mathcal{V})^{2}
-2DH_c(4\hspace{.1cm}\mathcal{U}+\mathcal{V})^{2}
-\frac{C}{3}(\mathcal{U}+\mathcal{V})(4\,\mathcal{U}+\mathcal{V})\hspace{.1cm}.\
\end{equation}
Now taking
$\mathcal{V}=h(\mathcal{U})=a\,\mathcal{U}^2+b\,\mathcal{U}^3+\cdots$,
the coefficients $a$ and $b$ can be found by using the chain rule
\begin{equation}\label{53}
\dot{\mathcal{V}}=h'(\mathcal{U})\dot{\mathcal{U}}\hspace{.1cm}.
\end{equation}
By this way, the problem effectively becomes one-dimensional.
Using eqs.(\ref{51}) and (\ref{52}), the coefficients of
$\mathcal{U}^2$-terms of eq.(\ref{53}) result in the parameter $a$
as follows:
\begin{equation}\label{54}
a=-\frac{1}{3H_c}\left(\frac{8}{3}+32H_cD+\frac{B}{18H_c}+\frac{4}{3}C\right)\hspace{.1cm}.
\end{equation}
The coefficient $b$ can be also found by the $\mathcal{U}^3$
terms. Using the expansion $\mathcal{V}=a\,\mathcal{U}^2+\dotsc$,
eq.(\ref{51}) leads to
\begin{equation}\begin{split}\label{55}
\dot{\mathcal{U}}&=\left(\frac{2}{3}+32H_cD+\frac{B}{18H_c}+\frac{4}{3}C\right)\mathcal{U}^2+
\left(\frac{4}{3}+16H_cD+\frac{B}{9H_c}+\frac{5}{3}C\right)a\hspace{.1cm}\mathcal{U}^3+\cdots\\
&=\alpha\hspace{.1cm}\mathcal{U}^2+\beta\hspace{.1cm}\mathcal{U}^3+\cdots\hspace{.1cm}.\
\end{split}\end{equation}
So for the cases with zero eigenvalues, the higher order terms,
through eq.(\ref{55}), must be considered in studying the
stability behavior of attractors. Note that the absence of the
linear terms in eq.(\ref{55}) reflects the fact that $\lambda_1$
is zero and we must focus on the next-leading terms. The attractor
is then stable if $\alpha<0$. For the cases where $\alpha=0$, we
must look at the sign $\beta$. $\beta<0$ leads to stable
attractors. If again $\beta$ becomes zero, we must go to higher
orders.

As an explicit example, we consider
\begin{equation}\label{56}
F(R,G)=R+\frac{1}{3}R^{3}-\frac{1}{3}\hspace{.1cm}.\
\end{equation}
The critical points are obtained by solving eqs.(\ref{20}) and
(\ref{21}), which result in
\begin{equation}\label{57}
(R_c,H_c)=\left(1,\frac{1}{\sqrt{12}}\right)\hspace{.1cm}.\
\end{equation}
Note that we have not considered the unphysical solution $R_c=-2$.
It can be easily seen that $\eta=0$ (using (\ref{29})), therefore
$\lambda_1=0$ and $\lambda_2=-3H$. Despite a negative eigenvalue,
the numerical calculations show that the critical point (\ref{57})
is not a stable attractor (see Fig.7)

This can be justified by calculating the parameter $\alpha$ and
$\beta$ of eq.(\ref{55}), which leads to
\begin{equation}\label{58}
\dot{\mathcal{U}}=\frac{4}{3}\hspace{.1cm}\mathcal{U}^2-\frac{100}{9\sqrt{3}}
\hspace{.1cm}\mathcal{U}^3
+\dotsc\hspace{.1cm}.\
\end{equation}
Since $\alpha>0$, the critical point is not stable, in accordance
with Fig.7.
\begin{figure}[hbt]\label{Figure 7}
\centering
\includegraphics*[height=5cm,width=5cm]{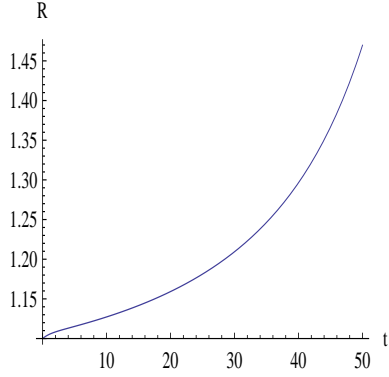}\hspace{1cm}
\caption{The plot of $R(t)$ of
$F(R,G)=R+\frac{1}{3}R^{3}-\frac{1}{3}$. The initial values are
$(R_0,H_0)=(1.1,0.3)$. It is clear that system does not approach
$R_c=1$.  }
\end{figure}
\section{Singular points of $F(R,G)$}
As pointed out after eq.(\ref{22}), in deriving the critical point
equation (\ref{20}), it has been assumed that the denominator of
eq.(\ref{17}) is finite at the critical values, so $\dot{R}=0$
leads us to set the numerator of eq.(\ref{17}) equal to zero. But,
as we will show, for the cases where the Lagrangian $F(R,G)$ has
some singularities, this assumption, i.e. the finiteness of the
denominator of eq.(\ref{17}), is not right and we must carefully
reinvestigate our results.

Take the function $F(R,G)$ as follows
\begin{equation}\label{59}
F(R,G)=\frac{P(R,G)}{Q(R,G)},\
\end{equation}
where $Q(R,G)$ has a root of order $n$ at $R=\alpha$, i.e.
\begin{equation}\label{60}
Q(R,G)=(R-\alpha)^{n}g(R,G).\
\end{equation}
It can be easily seen that in this case, the denominator of
eq.(\ref{17}) diverges at $R\rightarrow\alpha$, with the power
greater than numerator. Substituting eq.(\ref{59}) into
eq.(\ref{17}), results in
\begin{equation}\label{61}
\dot{R}=\frac{(6H^{2}-R)PQQ_{R}+8H^2(R-12H^2)^{2}(2PQ_{R}Q_{G}-PQQ_{RG}-QP_{G}Q_{R})+\cdots}
{6H(2Q_{R}^{2}-QQ_{RR})P+\cdots}\hspace{.2cm},
\end{equation}
where the dots denote the higher order terms of $R-\alpha$. Power
counting of eq.(\ref{61}) shows that
\begin{equation}\label{62}
\dot{R}=\frac{O((R-\alpha)^{2n-1})}{O((R-\alpha)^{2n-2})}\hspace{.2cm},\
\end{equation}
which results an extra solution for $\dot{R}=0$ as follows
\begin{equation}\label{63}
(R_c,H_c)=\left(\alpha,\sqrt{\frac{\alpha}{12}}\right)\hspace{.2cm}.
\end{equation}
$H_c$ is found from eq.(\ref{21}).

To study the stability of this critical point, we need to
calculate the eigenvalues of the matrix $M$ (in eq.(\ref{25})). A
lengthy calculation shows that
\begin{equation}\label{64}
\delta\dot{R}=-H\left[\frac{Q_{R}^{2}+QQ_{RR}}{2Q_{R}^2-QQ_{RR}}-\frac{QQ_{R}
(3Q_{R}Q_{RR}-QQ_{RRR})}{(2Q_{R}^2-QQ_{RR})^{2}}+\cdots\right]\delta
R\hspace{.2cm},\
\end{equation}
where using the expression (\ref{60}) for $Q(R,G)$, leads to
\begin{equation}\label{65}
\delta\dot{R}=-\frac{H_c}{n+1}\delta R+\cdots\hspace{.2cm},\
\end{equation}
at $R=\alpha$. So the matrix $M$ becomes
\begin{equation}\label{66}
M=\left(\begin{array}{llll} -4H_c&\hspace{.7cm}-1/6\\
0&-H_c/(n+1)
\end{array}\right),
\end{equation}
with eigenvalues
\begin{equation}\label{67}
\lambda_1=-4H_c\hspace{.5cm},\hspace{.5cm}\lambda_2=-\frac{H_c}{n+1}
\hspace{.2cm},\
\end{equation}
where both of them are real negative numbers. So we lead to an
important general consequence:
 {\it Any singularity of the function $F(R,G)$
is an stable attractor solution}.

The same is true for the cases where $Q(R,G)$ has a root of order
$n$ at $G=\beta$:
\begin{equation}\label{68}
Q(R,G)=(G-\beta)^ng(R,G) \hspace{.2cm}.\
\end{equation}
The same procedure results in a critical point at $G_c=\beta$, or
\begin{equation}\label{69}
(R_c,H_c)=\left(\sqrt{6\beta},\left(\frac{\beta}{24}\right)^{1/4}\right)\hspace{.2cm},\
\end{equation}
in which eq.(\ref{22}) has been used. The matrix $M$ becomes   the
same as eq.(\ref{66}), which proves that this critical point is a
stable attractor.

So generally for
\begin{equation}\label{70}
Q(R,G)=(R-\alpha)^n(G-\beta)^mg(R,G) \hspace{.2cm},\
\end{equation}
the model has the stable attractor points (\ref{63}) and
(\ref{69}).

As an example, we consider the model discussed in section 3-2,
that is $F(R,G)=R+Gf(R)$. The regular critical points
(non-singular type) can be found by solving the relation
(\ref{38}), and the stability condition is eq.(\ref{39}). Now
consider the explicit example
\begin{equation}\label{71}
F(R,G)=R+\frac{mG}{R^{2}-1} \hspace{.2cm},\
\end{equation}
where $m$ is a constant. Eq.(\ref{38}) leads to
\begin{equation}\label{72}
3(R^{2}-1)^{2}+mR^{3}=0 \Longrightarrow R_{c}=R_{c}(m)
\hspace{.2cm},\
\end{equation}
and the inequality (\ref{39}) results in
\begin{equation}\label{73}
\frac{2(R_c^2-1)}{R_c^2-5}>1 \hspace{.2cm}.\
\end{equation}
For $R_c^2-5>0$, eq.(\ref{73}) leads to $R_c^2>-3$, which is
always true, and for $R_c^2-5<0$, it results in $R_c^2<-3$, which
is never true. So the condition (\ref{73}) holds if
\begin{equation}\label{74}
R_c^2>5\hspace{.2cm}.\
\end{equation}
Now if we chose $m=-10$, the eq.(\ref{72}) gives two following
real solutions:
\begin{equation}\label{75}
R_{1c}=0.534\hspace{.5cm},\hspace{.5cm}R_{2c}=3.837\hspace{.2cm}.\
\end{equation}
It is clear that $R_{1c}$ does not satisfy (\ref{74}), while
$R_{2c}$ does. Explicit calculation of $\eta$ (in eq.(\ref{29}))
shows that $\eta|_{R=R_{1c}}<0$ and $\eta|_{R=R_{2c}}>0$. So we
expect that the stable critical point of
\begin{equation}\label{76}
F(R,G)=R-\frac{10G}{R^{2}-1} \hspace{.2cm},\
\end{equation}
is
\begin{equation}\label{77}
(R_c,H_c)=(3.837,0.565) \hspace{.2cm}.\
\end{equation}
Numerical calculation verifies this. See Fig.8
\begin{figure}[hbt]\label{Figure_8}
\centering
\includegraphics*[height=5cm,width=5cm]{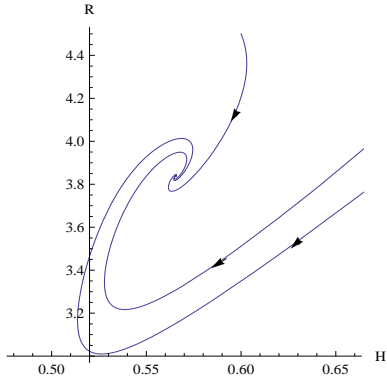}\hspace{1cm}
\caption{ The spiral paths leading to attractor (\ref{77}) of the
Lagrangian $F(R,G)=R-10G/(R^2-1)$.}
\end{figure}

Until now, we find the regular attractor of (\ref{76}). But it is
clear that the $F(R,G)$ in eq.(\ref{76}) is singular at $R=1$. So
we expect another stable attractor at the point
\begin{equation}\label{78}
(R_c,H_c)=\left(1,\sqrt{\frac{1}{12}}\right) \hspace{.2cm}.\
\end{equation}
This new attractor is also verified by the numerical method. See
Fig.9
\begin{figure}[hbt]\label{Figure_9}
\centering
\includegraphics*[height=5cm,width=5cm]{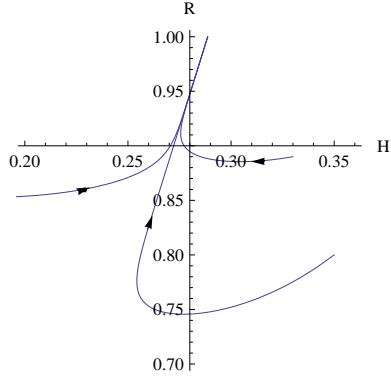}\hspace{1cm}
\caption{ The paths of $F(R,G)=R-10G/(R^2-1)$ leading to attractor
$(R_c,H_c)=\left(1,\sqrt{1/12}\right)$}
\end{figure}

\section{The critical curves }
There are other interesting cases in which the critical points are
replaced by critical curves. In this case, each of the infinite
points on this critical curve are in fact a critical point, and
besides, as we will show, they are stable attractors. This
situation occurs when the criticality condition (\ref{20}) holds
for any $R$ and $H$ values.

Before introducing some special examples, we first prove a general
statement:

If a $F(R,G)$ function satisfies (\ref{20}) and $R$ and $G$
satisfy eqs.(\ref{21}) and (\ref{22}), respectively, then $\eta =
{F_R}/(3A)-4H^2$ is equal to zero.\\
{\bf proof}: Since condition (\ref{20}) holds for any $R$ and $H$,
it can be differentiated with the result
\begin{equation}\label{79}
(\frac{1}{2}RF_{RR}-\frac{1}{2}F_{R}+GF_{RG})dR+(\frac{1}{2}RF_{RG}+GF_{GG})dG
= 0 .
\end{equation}
But from eq.(\ref{14}) we have
\begin{equation}\label{80}
dG=8H(R-12H^2)dH+4H^2dR,
\end{equation}
so
\begin{equation}\begin{split}\label{81}
[&\frac{1}{2}RF_{RR}-\frac{1}{2}F_R+
GF_{RG}+4H^2(\frac{1}{2}RF_{RG}+GF_{GG})]dR \\
&+8(\frac{1}{2}RF_{RG}+GF_{GG})H(R-12H^2)dH=0.
\end{split}\end{equation}
Using eq.(\ref{21}), the coefficient of $dH$ becomes zero. The
coefficient of $dR$, which now must be set to zero, specifies
$F_R$ as follows
\begin{equation}\label{82}
F_R=R(F_{RR}+8H^2F_{RG}+16H^4F_{GG})=RA,
\end{equation}
in which the function $A$, introduced in (\ref{27}), has been
used. Therefore $\eta$ becomes
\begin{equation}\label{83}
\eta=\frac{F_R}{3A}-4H^2=\frac{1}{3}R-4H^2=0 ,
\end{equation}
where $R=12H^2$ has been used. This completes our
proof.$\blacksquare$

Now we note that if $F(R,G)$ satisfies (\ref{20}), this equation
does not impose any extra constraint on $F(R,G)$ and therefore
does not specify any critical values for $R$ and $H$. So the only
remaining relation in the $R-H$ phase-space plane is the second
equation (\ref{21}), which defines a critical curve. The
eigenvalues of this critical curve, as a result of the above
statement which leads to eq.(\ref{83}), are :
\begin{equation}\label{84}
\lambda_1=0\hspace{.5cm},\hspace{.5cm}\lambda_2=-3H.\
\end{equation}
But it can be shown that in the case of the emergence of critical
curve, the stability of any particular point on this curve can be
determined by the nonzero eigenvalues~\cite{bar}. Since in our
case, $\lambda_2=-3H<0$, therefore {\it any points on the critical
curve $R=12H^2$ of $F(R,G)$ models is a stable attractor}. This is
a general result.

Now let us consider some explicit examples.

{\bf Example 1}: Let us first consider the class of models
introduced in section 3-1. For $F(R,G)=F(RG)$, it is obtained that
eq.(\ref{20}) results in
\begin{equation}\label{85}
\frac{3}{2}xF'(x)-F(x)=0,\
\end{equation}
where $x=RG$. If we demand that the eq.(\ref{85}) satisfies for
all $x$, then it can be viewed as a differential equation with
solution $F(x)=x^{2/3}$. So the $F(R,G)$ model
\begin{equation}\label{86}
F(R,G)=(RG)^{2/3}\
\end{equation}
has a critical curve $R=12H^2$. All the points on this curve are
stable attractors. Figs.(10) and (11) show that both
$H_{1c}=0.977$ and $H_{2c}=1.99$ points (as two arbitrary points),
with $R_{1c}=11.454$ and $R_{2c}=47.52$, respectively,  are stable
attractors of this model.
\begin{figure}[hbt]\label{Figure_10}
\centering
\includegraphics*[height=5cm,width=5cm]{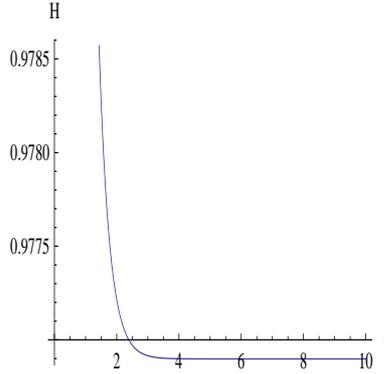}\hspace{1cm}
\caption{ The plot of $H(t)$  for $F(R,G)=(RG)^{2/3}$ model. The
point $(H_c,R_c)=(0.977,11.454)$ is a stable attractor.}
\end{figure}

\begin{figure}[hbt]\label{Figure_11}
\centering
\includegraphics*[height=5cm,width=5cm]{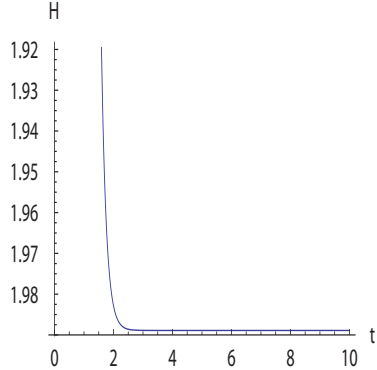}\hspace{1cm}
\caption{ The plot of $H(t)$  for $F(R,G)=(RG)^{2/3}$ model. The
point $(H_c,R_c)=(1.99,47.52)$ is a stable attractor.}
\end{figure}
{\bf Example 2}: consider the following $F(R,G)$ model:
\begin{equation}\label{87}
F(R,G)=R^nf(G^kR^m)\
\end{equation}
with arbitrary constants $n,k$ and $m$. Substituting (\ref{87})
into eq.(\ref{20}), results in:
\begin{equation}\label{88}
(\frac{n}{2}-1)R^nf(x)+(\frac{m}{2}+k)R^{n+m}G^{k}f'(x)=0
\end{equation}
where $x=G^kR^m$. If one demands the above equation satisfies for
all $R$ and $G$s and for any arbitrary function $f(x)$, then the
constants $n,k$ and $m$ must satisfy
\begin{equation}\label{89}
n=2\hspace{.5cm},\hspace{.5cm}m=-2k .\
\end{equation}
So $F(R,G)=R^2f(G^k/R^{2k})=R^2g(G/R^2)$ satisfies (\ref{20}) and
the curve $R=12H^2$ is its critical curve. It is interesting to
note that the case considered in example 1, i.e. the
eq.(\ref{86}), is in fact of this type:
$(RG)^{2/3}=R^2(G/R^2)^{2/3}$.

{\bf Example 3}: Consider the following model
\begin{equation}\label{90}
F(R,G)=\alpha G+f(R)\hspace{.1cm},\
\end{equation}
then eq.(\ref{20}) results in:
\begin{equation}\label{91}
Rf'=2f \
\end{equation}
which its solution, as a differential equation, is $f(R)=\beta
R^2$. So all models of the type
\begin{equation}\label{92}
F(R,G)=\alpha G+\beta R^{2}\hspace{.1cm},\
\end{equation}
have $R=12H^2$ as their critical curve, with infinite number of
stable attractors.

This example also shows that in the case of $f(R)$ gravity
theories, which is the $\alpha=0$ case of eq.(\ref{90}), the only
model which leads to the critical curve $R=12H^2$ is the
$f(R)=R^2$ model.

{\bf Example 4}: As the last example, consider the model
\begin{equation}\label{93}
F(R,G)=R+Gf(R).
\end{equation}
Substituting (\ref{93}) into eq.(\ref{20}), results in
\begin{equation}\label{94}
R^2f'(R)=6\Rightarrow f(R)=-\frac{6}{R}.
\end{equation}
So $F(R,G)=R-6G/R$ has also the critical curve $R=12H^2$.

The above mentioned procedure can be applied to some other
functional forms, such as $F(R,G)=f(G/R)$, with the result
$F=(G/R)^{2}$, etc.

\section{Conclusion}
As a candidate of dark energy, we consider the generalized
Gauss-Bonnet dark energy models looking for the situations where
the late-time behavior of this modified gravity theory is the
de-Sitter space-time. We describe the phase space of this theory
by the two-dimensional $R-H$ space. This dimensionality verifies
by the fact if the three-dimensional $R-H-G$ space has been
chosen, one of the eigenvalues of stability matrix is always zero,
which indicates that the number of independent variables is two.

The eigenvalues of stability matrix show that the critical points,
i.e. the de-Sitter space-times, are, in general, the stable
attractor if $\eta = {F_R}/({3A})-4H^2>0$, a fact that has been
verified by several examples. The emergence of critical points
with $\eta =0$, in which one of the eigenvalues is zero,
$\lambda_1=0$, forces us to consider the higher order terms in
normal basis in order to have a correct judgment about the
stability of these kinds of critical points.

We also find two classes of stable attractors: the singular points
of the Lagrangian $F(R,G)$ and the cases where the critical points
are replaced by the critical curve $R=12H^2$ (in $R-H$ plane). In
the latter case, all the points on this curve are stable
attractors.

{\bf Acknowledgement:} This work was partially supported by the
"center of excellence in structure of matter" of the Department of
Physics of the University of Tehran, and also a research grant
from the University of Tehran\\ \\

\end{document}